%% file: ictel2010.tex
\documentclass[a4paper,twoside]{article}

\usepackage{epsfig}
\usepackage{subfigure}
\usepackage{calc}
\usepackage{amssymb}
\usepackage{amstext}
\usepackage{amsmath}
\usepackage{multicol}
\usepackage{pslatex}
\usepackage{apalike}
\usepackage{fancyhdr}
\usepackage{insticc}
\usepackage[small]{caption}

\begin{document}
\onecolumn \maketitle \normalsize \vfill

\input{introduction}

\input{background}

\input{strengths}

\input{survey}

\section{\uppercase{Conclusions}}
\label{sec:conclusion}

\noindent Computer-based learning methodologies and related infrastructures are essential nowadays at any level of the academic world. Computer labs and required ICT are becoming an economic and staff resource barrier for all educational institutions. Thus efforts to provide more effective solutions will have a practical impact on education.

Thin client technologies have proven to be a good option for reducing budget and maintenance requirements in scenarios with more than five to eight computers. However, it would be a mistake to defend this solution without examining what use is going to be made of the lab equipment. There are many educational scenarios where thin clients are cheaper to use, easier to maintain and easier to upgrade. In this paper, we have identified the main advantages and limitations of thin clients regarding the most common scenarios. A widespread of this technology in some cases would improve global computer literacy.

%\vfill

\section*{\uppercase{Acknowledgements}}

\noindent We would like to thank the Universidad Polit\'ecnica de Madrid's Office of Cooperation for the Development for its financial an logistic support\footnote{This work is partially supported by the project MESTUN (Monitoring, evaluation and technological sustainability of the University of Ngozi) funded by the call 2009 of the Direction of Cooperation for the Development of the Technical University of Madrid.}, and the members of TEDECO group for their interest, constructive comments and active participation in the development of this paper.

\renewcommand{\baselinestretch}{0.98}
\bibliographystyle{apalike}
{\small
\bibliography{ictel2010}}
\renewcommand{\baselinestretch}{1}

\end{document}

%% file: introduction.tex
\section{\uppercase{Introduction}}
\label{sec:introduction}

\noindent Traditional education has been hampered by the difficulties inherent in information and communication technologies (ICT). Nowadays, modern education faces a new challenge: the use of technology in the classroom. Concerning these new educational challenges, probably one of the most popular is related to improving computer lab performance. Note that considering the number of institutions that have to deal with these infrastructures, any advantageous proposal will have a worldwide repercussion.

There are a set of well-known problems that educational institutions must address when using computer labs. The main obstacles are usually related to economic issues. Some institutions do not even have computer labs due to unaffordable costs. The equipment and the installation of a new computer lab is very expensive. Moreover, both software and hardware maintenance is a sustained cost over time. Software and hardware evolve daily at a speed that most institutions are unable to keep pace with. Thus software and hardware upgrades especially are an expense that many institutions are unwilling to pay.

Thin client technology can be used to overcome some of these problems. For instance, this technology is an especially interesting option when upgrading an educational computer lab, but it is not widespread in these scenarios probably due to historical prejudices. An excessive triumphalism by supporters of thin client technology, claiming the many advantages of this technology sometimes rightly or wrongly but usually without proper justification, may have led to the early dismissal of thin client technology by many professionals. It has often been relegated to an immature technology status. However, today there are solutions based on thin client technology that are simple to use, and after many years of development and evolution are highly reliable, which could be a very interesting option in some specific scenarios. Some of these solutions are the \textit{Linux Terminal Server Project} (LTSP) project\footnote{http://www.ltsp.org}, especially designed for Unix-based systems, or \textit{Lan Core}\footnote{http://lancore.sourceforge.net}, a project originally focused on the use of thin clients on Windows systems.

In this paper we summarise the main characteristics of thin clients (Section~\ref{sec:background}), explicitly stating their strengths and weakness related to the educational environment (Section~\ref{sec:discussion}). From this information we have identified a set of scenarios (Section~\ref{sec:survey}) where the use thin client technology is a good option, and some counter examples.

%% file: background.tex
\section{\uppercase{Thin Client Technology}}
\label{sec:background}

\begin{figure}[t]
  \centering
   {\epsfig{file = 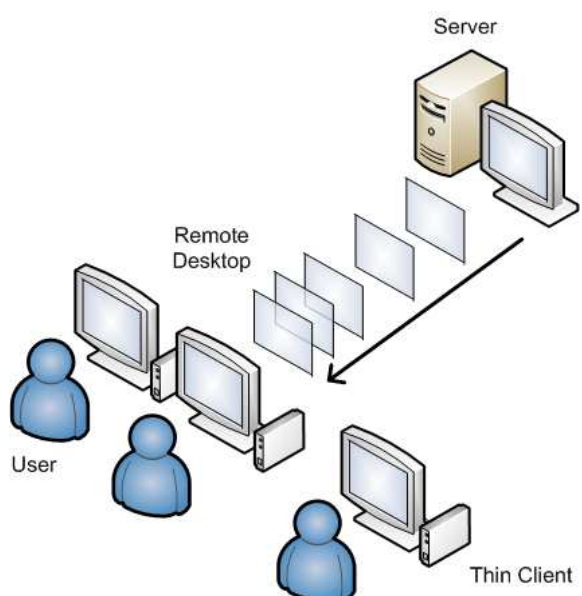, width =\columnwidth}}
  \caption{In a thin client network, all clients are continuously connected to a common server in a centralised architecture. For each connection the server sends the desktop to the client computer (the thin client). The user works using a remote desktop session.}
  \label{fig:remote-desktop}
\end{figure}

\noindent From the very beginning of computing, around the 1950s, early computers had to be shared by multiple users on economic grounds. Mainframes, mainly known for their large size and priced according by, evolved towards the interaction of a single computer with multiple users. A few years later the first terminal appeared: This was an specific device connected to the mainframe enabling multiple users to access and share computing resources through batch processing. The early terminals, today known as consoles, worked in text mode only, and were connected via serial lines. Therefore, a primitive type of general-purpose device working much like a thin client was there way back when the main commercial computers first came on the scene.

Console terminals evolved gradually and continuously for years until the advent of the personal computer (PC) in the 1980s. Physical terminals were then immediately replaced by PCs equipped with terminal emulation software. Not only were PCs a substitute for the old terminal devices, but they also marked a profound change in the computing architecture. For the first time, the client (i.e. the PC) had its own processing capacity, leading to the client and server division. It was also a turning point for terminal computing, where terminals became known as thin clients and they were no longer connected to a central computer directly, but via a network. New graphical operating systems, including the new Unix-based systems (e.g. Linux), introduced the new desktop concept. The client side in a thin client solution had to be modified to be adapted to graphical devices (see Figure~\ref{fig:remote-desktop}), new user interfaces, the mouse and other new input and output devices.

Because of everything discussed above, today it is not possible to find a primary reference introducing the new concept of thin client, terminal, or diskless devices in computing. However, there can be no doubt that thin client technology has been always present in computing, sometimes going by a different name or terminology, running on primitive or highly evolved hardware, even supported by entirely different operating systems, using batch processing, multitasking, and other hardware and software sharing strategies.

\subsection{Thin Client Benefits}

\begin{figure}[t]
  \centering
   {\epsfig{file = 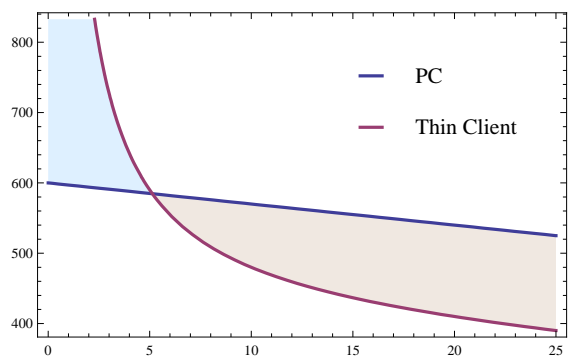, width =\columnwidth}}
  \caption{Evolution of the price (vertical axis in euros) per unit (PC or thin client) according to the number of installed workstations (horizontal axis). This estimate has been calculated assuming the following costs: PC 600 EUR, thin client 400 EUR, and server 1000 EUR; with an incremental discount ($n \times 0.5\%$) depending on the number of units purchased.}
  \label{fig:budget}
\end{figure}

\begin{figure}[t]
  \centering
   {\epsfig{file = 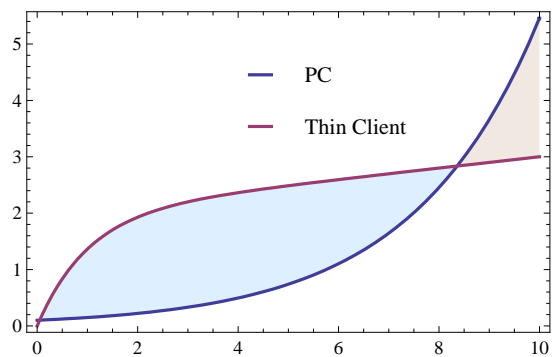, width =\columnwidth}}
  \caption{Estimated hours of maintenance per week (vertical axis) depending on the number of workstations (horizontal axis), comparing PCs versus thin clients.}
  \label{fig:maintenance}
\end{figure}

\noindent Thin client technology can make a network administrator's job easier for three main reasons: it is easy to manage, easy to secure and easy on the budget~\cite{Romm06}. An explanation of each one of these benefits follows:

\begin{enumerate}

\item A thin client network is easy to manage and maintain thanks to its centralised set-up and support. It simplifies troubleshooting, network management, and monitoring among other things.

\item Security is provided intrinsically by the architecture of a thin client solution. Input devices such as local Universal Serial Bus (USB) ports are easily disabled, thus avoiding a common source of viruses. The integration of firewall functionality in every computer in the network (i.e. a distributed firewall) is possible by just configuring a firewall on the server \cite{Ioannidis00}.

\item The use of thin client technology is low cost in terms of hardware (see Figure~\ref{fig:budget}), software (just one licence for the server), set-up and maintenance (see Figure~\ref{fig:maintenance}).

\end{enumerate}

Points one and two are related to the fact that there is only one processor and point three is related to the simplicity of the thin client technology. In the context of this paper there is one property of thin client technology that should be emphasised: a thin client has no need of high processing capacity. Note that it will not run anything apart from the hardware and software needed to properly visualise the graphical user interface. Accordingly low-performance hardware (i.e. out-of-date computers or ``ancient hardware'') could be used to build a thin client. Considering these benefits, thin clients are an attractive technology for institutions with old computers, budget concern and limited maintenance staff (i.e. more than 80\% of the educational institutions in the world, most of them in developing countries).

\subsection{Green Computing}

Reducing energy consumption is an attractive goal not only due to increasing energy costs, but also from the standpoint of natural resources and environmental conservation. There are many reasons for implementing green IT systems, and hardware that is specifically designed for thin clients seems to fit in perfectly here \cite{Fraunhofer08}.

In a comparison with other alternatives that have similar possibilities to thin clients on other counts, thin client technology has an important point in favour of its aptness for green computing.

%% file: strengths.tex
\section{\uppercase{Strengths and Weaknesses}}
\label{sec:discussion}

Probably the main feature of a thin client is that users do not appreciate any difference from a desktop PC when they first come into contact with the device. The desktop displayed by a thin client is exactly the same as one shown by a PC. This property of thin client technology is essential in educational labs and it is one of the strengths of this technology. Using thin clients students can acquire the same skills that they could learn using PCs because they actually use the same tools and applications (running on the server instead of on their own machines).

Another important strength of thin clients is that most of the applications used on a PC, such as those included in an office suite (e.g. Microsoft Office in Windows operating systems or Open Office for Linux) will probably run on a network with thin clients. Similarly, Wikipedia (probably the most visited website on the Internet in academic environments) also has these advantages.

On the other hand, the thin client technology's main weakness materialises in what are technically referred to as high-motion scenarios \cite{Deboosere07}. High-motion scenarios include:

\begin{itemize}

\item Streaming multimedia: video and audio are almost prohibitive using thin clients.

\item Motion graphics: animations, slide effects, etc., are processes that are often affected by a desktop virtualisation.

\item Applications requiring a powerful graphics card: everything related to games and 3D rendering.

\item Latency-sensitive applications, where the time it takes an event to reach the client through a network is critical. %not admissible.

\end{itemize}

Hence, thin client technology is not an alternative for scenarios with these requirements, but it could be an optimal solution in the near future due to  several technical developments attempting to overcome the above weaknesses \cite{Simoens08}.

%% file: survey.tex
\section{\uppercase{Common Academic Scenarios}}
\label{sec:survey}

\noindent Each computer-based environment has particular requirements that need to be considered to find the optimal hardware and software solution. Looking at thin clients, some advantages and limitations of this technology will act as detriment or incentive for a thin client solution. Below we analyse the main academic scenarios regarding thin client suitability.

Note that we have highlighted overlaps among different scenarios.

\subsection{Computer Labs}
\label{sec:computer-labs}

As discussed above (Figures~\ref{fig:budget} and~\ref{fig:maintenance}) thin clients have an edge in terms of budget and maintenance concerns over PCs in installations from five and eight computers respectively. This covers almost all computer labs in education. Thus a computer lab intended for traditional computer-based learning may also benefit from thin clients if they support the required tools. A previous study suggests that most tools used in the academic environment can be used with thin clients \cite{Brinkley06}. The results of this research are conclusive: of the 45 tools used in this case, only three could not be used, and there were some minor problems with another three. The rest of tools run correctly using thin client technology.

Technology facilitates student participation and computer labs are beginning to be used with this aim \cite{Peiper04}. Although there are interactivity increments in collaborative environments they are supported by thin client technology.

\subsection{Distance Learning}
\label{sec:distance-learning}

Distance learning is a relatively modern academic situation in computer-based learning, where ICTs are used to circumvent the boundaries of a physical classroom. This is a wider scenario than the one considered in a computer lab. A range of situations depending on student and professor location are listed below:

\begin{itemize}

\item Local students and professors. Traditional computer-enhanced learning to meet today's ICT literacy need. This situation is spreading fast and is being combined with novel trends, such as digital ink (Section~\ref{sec:digital-ink}), which are suitable to work on thin clients.

\item Local students with a remote professor. Students gathered in a computer lab (see Section~\ref{sec:computer-labs}) attend a lecture by a remote professor. Not a common situation at traditional universities, it is an optimal solution in most cooperation for development scenarios. It is also suitable for thin client implementations due to the number of computers required and the limited budget and staff available.

\item Remote students with a local professor. Students using off-campus computers to access teaching resources. This is the new model being adopted by universities worldwide and not suitable for thin clients \cite{Casella07}.

\item Remote students and professors. Pure e-learning scenario also not suitable for thin clients.

\end{itemize}

Although the last two situations rule out a thin client solution given that the academic organization is not providing computers to access the resources, thin client implementations have been proven to be a feasible solution for use in most ``local student'' situations.

\subsection{Web-based Learning}
\label{sec:web-based-learning}

Web-based solutions are in widespread use for e-learning. These solutions could be part of distance learning or e-learning generally, but we have preferred to examine them separately due to their wide dissemination.

Special-purpose hardware for thin clients or common PCs operate here as thin clients for remote learning, since its limitations do not affect most web applications. Some educational solutions that use web-based services have already proposed the use of thin client technology in the classroom \cite{Andria07}. Even web-based scenarios including complex requirements are good scenarios for the use of thin clients \cite{Callaghan07}, provide they do not use multimedia or high-motion services.

\subsection{Digital Ink}
\label{sec:digital-ink}

New technologies are making a claim for a place in education in the classroom. Most of these technologies are publicised as new e-learning strategies, but some, such as digital ink, should be studied separately. It is not possible to decide whether a scenario is suitable for the use of thin client technology, based only on the above strengths and weaknesses. In this case, digital ink should be singled out from other e-learning strategies because it proposes the use of special-purpose hardware devices instead of common computers used in conventional e-learning. For example, a proposal for the use of the \textit{Classroom Presenter}\footnote{http://classroompresenter.cs.washington.edu} application (a software platform for digital ink) recommends the use of tablet PCs by students. Table PCs are really modern devices for which a thin client is no suitable \cite{Anderson07}.

\subsection{Education in Developing Countries}
\label{sec:developing-countries}

Health and education are fundamental aspects for a country's development. Education, especially in technical fields, is an important goal if developing countries are to emulate the information access opportunities of developed nations \cite{Krikke04}. However, an educational institution in a developing country cannot afford the cost of setting up and maintaining a computer lab. Cooperation for development guidelines call for the use of sustainable technologies and solutions, meaning the purely financial contribution of purchasing a computer lab is not acceptable. Thin client technology could further the use of computer labs in developing countries in a sustainable way. In this regard the following is an example of a success story in terms of the technology proposed in this paper.

In 2006 the Technology for Development and Cooperation (TEDECO) Group\footnote{http://tedeco.fi.upm.es} led by professors of the School of Computing at Universidad Polit\'ecnica de Madrid, Spain, launched a new distance education project at University of Ngozi (UNG), in Burundi. A committee formed by members of the TEDECO group and expert engineers explored several options for setting up two computer labs at UNG. The limited budget and the availability of very old computer resources were key points to consider when making a final decision \cite{Martinez09}.

The use of thin client technology in computer labs proved to be a success, especially taking into account that it was tested in a really tough environment, where there were no qualified staff, apart from a technical support staff working exclusively on maintenance. The project was considered a success due to \cite{Martinez10}: (i) the increase in the number of subjects that were offered over an e-learning platform like Moodle using the new computer labs, (ii) the increase in the number of students choosing to study at UNG, and (iii) the offer of a new course (fourth year) as result of student demand and the new subjects offered.

\subsection{Unsuitable Scenarios}

Although tt is important to ascertain the main scenarios where the use of thin client technology is suitable, it is even more necessary to identify scenarios where this technology should be avoided. To identify scenarios that are unsuitable for the use of client technology, always keep in mind the key weakness highlighted in Section~\ref{sec:discussion}: high-motion. A list of scenarios that are not suitable for thin clients follows:

\begin{itemize}

\item Distance learning, unless it requires the physical presence of students in a classroom.

\item Multimedia learning, which is a discipline that has been especially developed in certain sectors of education, such as medicine \cite{Ruiz06}.

\item Mobile learning. As stated in Section~\ref{sec:digital-ink}, e-learning strategies calling for student mobility can not be implemented with thin clients. These scenarios require the use of special devices that are not covered by the current thin client technology.

\end{itemize}

Although there are unsuitable scenarios related to multimedia, video, and other high-motion resources, the majority of computer-based scenarios do not require these features. Therefore, thin client technology is an alternative that is worthwhile considering.